\documentclass[aps,pre,twocolumn,floatfix]{revtex4-1}

\usepackage{amsmath}  
\usepackage{amsfonts} 
\usepackage{graphicx} 
\usepackage{epstopdf}
\usepackage{gensymb}
\usepackage{color}

\begin{document}

\title{Temperature Dependence of Nylon and PTFE Triboelectrification}

\author{Isaac A. Harris, Melody X. Lim, Heinrich M. Jaeger}
\affiliation{Department of Physics and James Franck Institute, The University of Chicago, 5720 S. Ellis Ave, Chicago, Illinois 60637, USA}

\begin{abstract}
Contact electrification, or tribocharging, is pertinent to a broad range of industrial and natural processes involving dielectric materials. However, the basic mechanism by which charge is transferred between insulators is still unclear. Here, we use a simple apparatus that brings two macroscopic surfaces into repeated contact and measures the charge on the surfaces after each contact. We vary the temperature of the surfaces, and find that increasing temperature leads to a decrease in the magnitude of charge transfer. When paired with a Monte-Carlo simulation and TGA measurements, our results support a mechanism where adsorbed surface water is crucial for charge exchange. Our setup is easily extendable to a variety of industrially relevant materials.
\end{abstract}

\maketitle

\section{Introduction}

Contact charging forms the basis for many industrial applications, including powder coating~\cite{bailey1998science}, cleaning surfaces~\cite{kawamoto2015electrostatic}, and air filtration~\cite{jaworek2007modern}. At the same time, tribocharging also presents a challenge to powder handling in industrial processes, as the separation of charge may cause the formation of clusters~\cite{ciborowski1962electrostatic,boland1969static,liang1996electrostatic,cocco2010particle}, or create sparks that damage electronic components~\cite{voldman1999state}. In addition to its widespread industrial relevance, the buildup of charge from frictional collisions is thought to be responsible for natural phenomena ranging from lightning inside volcanic ash clouds~\cite{brook1974lightning,pahtz2010particle} to potentially also the very early stages of the formation of planetesimals from interstellar dust~\cite{blum2008growth}. 

Despite the importance of phenomena related to triboelectric charging, the underlying mechanism for charge exchange is not well understood. In particular, it is unclear how insulators, which have very low charge mobility, can transfer charge. Several mechanisms for tribocharging have been proposed, including the transfer of electrons in trapped states~\cite{lowell1986triboelectrification,liu2010electrostatic,lacks2008nonequilibrium}, ion transfer~\cite{mccarty2008electrostatic,lee2018collisional}, and the transfer of nanoscale pieces of charged material~\cite{baytekin2011mosaic}.

Tribocharging has been shown to be sensitive to a wide variety of conditions, including the ambient environment conditions~\cite{greason1999investigation,matsuyama1995charge,schella2017influence}, impact velocity~\cite{xie2016effect}, material strain~\cite{sow2012strain}, the contact force~\cite{liu2013contact}, and the microscopic details of the surface contact~\cite{williams2012triboelectric,horn1992contact}. Previous work has reduced these effects by making highly controlled, precision measurements of the impact charging of a small particle~\cite{matsuyama2003impact,watanabe2006measurement,lee2018collisional,haeberle2018double}, or by using single-crystal materials that are flat to nanometer precision~\cite{stern1988deposition,collins2018simultaneous}. However, even with well-controlled surfaces, the inherent stochasticity of tribocharging can lead to a large variance in the results. 

We present an experiment that takes a different approach by measuring the average charge transfer between macroscopic materials. In doing so, we effectively average over many microscopic charge transfer processes.  Despite the fact that we do not control all microscopic aspects of the surface contact, we show that we are nevertheless able to extract robust statistical trends from our data. Here, we measure the charge transfer between materials on opposite ends of the triboelectric series~\cite{mccarty2008electrostatic}: nylon and polytetrafluouroethylene (PTFE). As we increase the surface temperature of the macroscopic samples, we observe a decrease in the overall magnitude of charge transfer. Our data, in conjunction with simulations and thermogravimetric analysis (TGA) measurements, suggest that adsorbed surface water plays a key role in the charge transfer between dielectric surfaces. 

\section{Experimental methods and results}
\label{sec:experiment}

\begin{figure}[h!]
\centering
\includegraphics[width = 0.97\columnwidth]{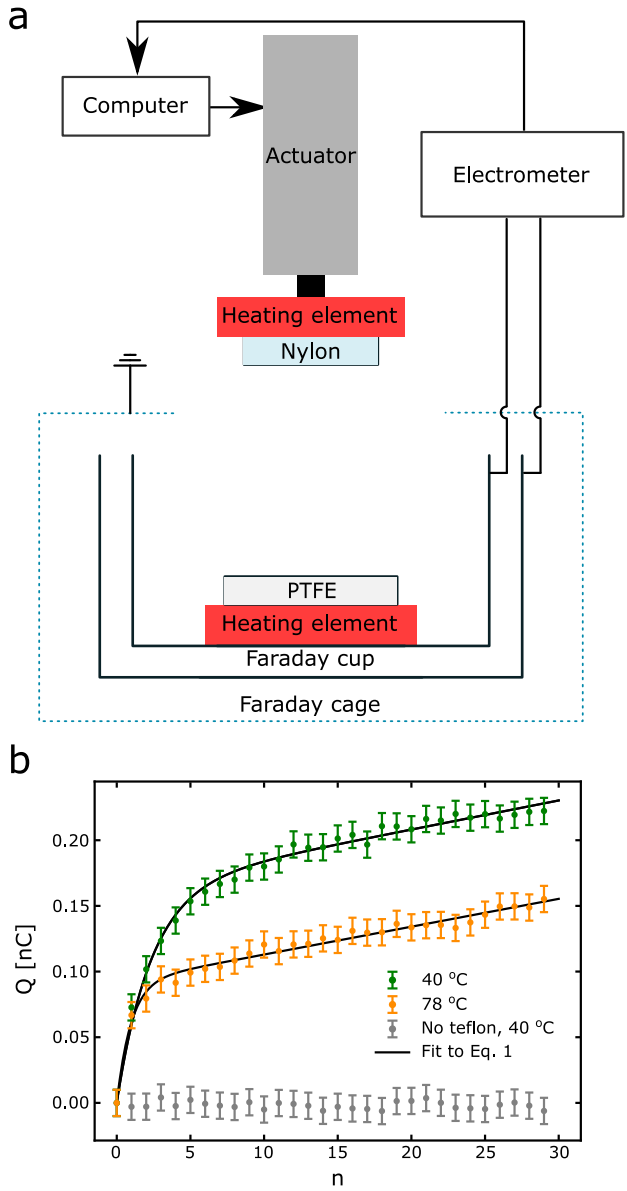}
\caption{a) Schematic of experimental setup. A piece of nylon is attached to a PID-controlled heating element. Both the nylon and the heating element are attached to the end of a linear actuator, which is computer-controlled to lower the nylon into a Faraday cup inside a Faraday cage, where it makes contact with a piece of PTFE. Another PID-controlled heating element is attached to the PTFE and to the base of the Faraday cup. The charge on the nylon is measured before and after the nylon touches the PTFE using an electrometer. b) Example time-series of the charge on the nylon. Over repeated contacts, the nylon acquires a positive charge. Curves in green (orange) show data taken with the plastics at 40$^\circ$C (78$^\circ$ C). Data is normalised so that the initial charge is 0nC, and then fit to an exponential with linear term. Data taken without the PTFE at 40$^\circ$ C are shown in grey. Error bars indicate the uncertainty in the electrometer reading during a collision.}
\label{fig:Apparatus}
\end{figure}

Our setup is illustrated in Fig.~\ref{fig:Apparatus}a. A macroscopic piece of PTFE ($l\times w\times d=$~30$\times$30$\times$3.2 mm$^3$) is housed in a Faraday cup, which is shielded by a Faraday cage. At the same time, a matching piece of nylon ($l\times w\times d=$~25$\times$25$\times$3.2 mm$^3$) is attached to the end of a pneumatic actuator, which lowers the nylon into the Faraday cup, making contact with the PTFE. Both the nylon and PTFE are cleaned before an experiment (see Appendix A for details). Our experiments take place in a humidity-controlled (42-44\% RH) environment. An electrometer (Keithley 6514) measures the total charge on nylon and PTFE during contact (10 seconds), then the charge on the PTFE after contact (10 seconds). These two measurements are then subtracted, such that the final measurement is the charge on the nylon after each collision. The apparatus thus allows for the automated control of a sequence of collisions, and the measurement of the charge as a function of collision number. 

In addition to automating repeated contact between a pair of dielectrics, our apparatus also allows for the independent control of the surface temperature of both the nylon and the PTFE during an experiment. A PID-controlled heating element and thermistor were placed on the back side of each dielectric (see Appendix B for calibration curves). During an experiment, the nylon and PTFE are both held at a specified temperature for 15 minutes, then undergo a sequence of 30 collisions over a time span of about 10 minutes. We vary the surface temperature of the dielectrics from room temperature (25$^\circ$C) to 80$^\circ$C. 

Representative examples of charging curves are shown in Fig.~\ref{fig:Apparatus}b. When the PTFE is removed from the Faraday cup (so that the nylon does not touch anything when the linear actuator moves downward) the charge on the nylon does not systematically change at any temperature (grey data), suggesting that over the timescales of the experiment (30 collisions), the dissipation of charge from the nylon is negligible. In contrast, when the nylon and PTFE are brought into repeated contact, the nylon acquires a positive charge that increases with collision number (green and orange data). This is consistent with previous measurements of the triboelectric series~\cite{clint2001acid,diaz2004semi,mccarty2008electrostatic,lacks2011contact,galembeck2014friction}. Qualitatively, the asymptotic accumulation of charge on the nylon at 78$^\circ$C is smaller than at 40$^\circ$C. These results are consistent with the trends reported in Ref.~\cite{greason1999investigation}.

\begin{figure}
\centering
\includegraphics[width = \columnwidth]{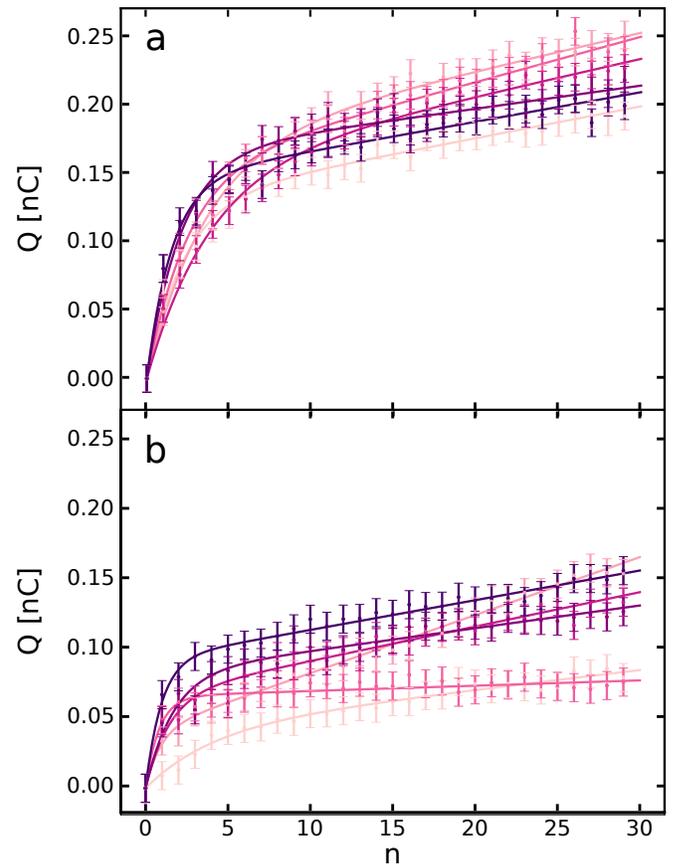}
\caption{Stochastic charging experiments on macroscopic materials yield statistically significant trends. Experimental time-series for the charge on the nylon, with surface temperature in a) 34~$^\circ$C, and in b) 77~$^\circ$C. Different colours represent different experiments. The data has been normalised so that the initial charge is 0nC, and then fit to Eq.~(\ref{eq:qn}) (solid lines). Error bars indicate the uncertainty in the electrometer reading during a collision.}
\label{fig:alldata}
\end{figure}

These trends are reproducible over many experiments, despite the relatively uncontrolled nature of the contact between the nylon and PTFE. Fig.~\ref{fig:alldata} shows examples of the charge time-series obtained for several different experiments, each with freshly prepared samples, at two different temperatures (34~$^\circ$C in part a, and 77~$^\circ$C for part b). There is variation between each time-series, due to microscopic differences between the surfaces. At the same time, the overall shape of the curves is highly consistent for each temperature. In light of the consistency in the data, we focus on average measures for each temperature, thus providing insight to the underlying statistical processes during contact charging.

Our data, as shown in Figs.~\ref{fig:Apparatus}b and~\ref{fig:alldata} exhibit a rapid increase of charge as a function of collision number, a behavior which has been reported in other contact electrification experiments as well~\cite{collins2018simultaneous,liu2013contact,greason1999investigation,apodaca2010contact, matsusaka2010triboelectric}. In addition to the initial increase, we typically observe a much more slowly increasing background, which similar experiments by others did not report.  We make these observations quantitative by fitting our experimental data to a variation of the condenser model proposed in Refs.~\cite{matsusaka2000electrification,matsusaka2007control,matsusaka2010triboelectric}. This phenomenological model provides a mechanism for an exponential buildup of charge towards some saturation value $q_\infty$. Although the model also includes a second exponential term to represent charge dissipation, our results (Fig.~\ref{fig:Apparatus}b) suggest that this term is not relevant within the timescales of our experiment.  In addition, we include a linear term to account for the slower background that we observe (as we will discuss below, we link this background to slow drift in the contacting portions of the surfaces being pushed together by the actuator). We thus fit the charge~$Q(n)$ as a function of collision number~$n$ to the following function: 

\begin{equation}
    Q(n) = q_{\infty}(1-e^{-n/n_0}) + An
    \label{eq:qn}
\end{equation} 
where~$q_\infty$,~$n_0$, and~$A$ are fitting parameters.  

\begin{figure}
\centering
\includegraphics[width = 0.95\columnwidth]{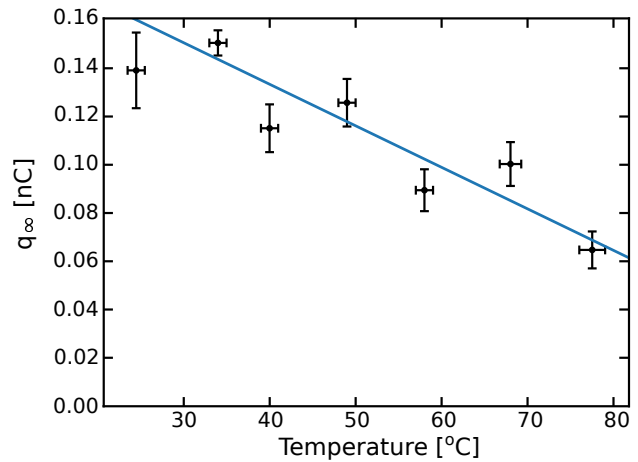}
\caption{Effect of temperature on~$q_\infty$ (see Eq.~(\ref{eq:qn}) for definition). As the temperature is increased,~$q_\infty$ decreases. Error bars denote the standard error in the mean of several measurements. We fit the data to a linear function of temperature~$a+bT$ (blue line), and find~$a = 0.2$nC,~$b=-1.7\times10^{-3}$nC/$^\circ$C. }
\label{fig:qSumz}
\end{figure}

Figure~\ref{fig:qSumz} shows the variation of~$q_\infty$ with surface temperature. Our surfaces are macroscopic and thus may vary widely in their microscopic properties, leading to previously observed stochastic charging processes~\cite{medley1950frictional,horn1992contact,matsusaka2010triboelectric,apodaca2010contact,haeberle2018double}. Despite this statistical nature of contact charging, we nevertheless observe an average decrease of~$q_\infty$ with increasing temperature. Given that~$q_\infty$ corresponds to the asymptotic saturation of charge on the surface in the absence of any other effects, Fig.~\ref{fig:qSumz} suggests that increasing the surface temperature reduces the capacity of the surfaces to exchange charge. This result suggests that contact charging in our experiment is mediated by a mechanism that is responsive to this small temperature change (room temperature to 80$^\circ$C).

\begin{figure}
\centering
\includegraphics[width = \columnwidth]{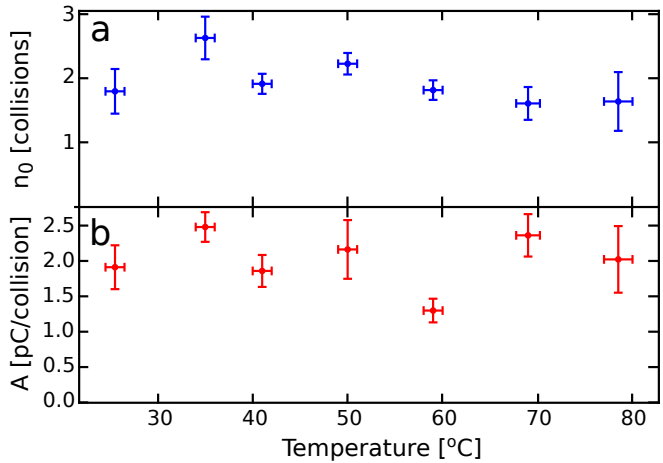}
\caption{Effect of temperature on (a)~$n_0$ and (b)~$A$. In contrast to~$q_\infty$, neither~$n_0$ nor~$A$ show systematic variation with temperature within the error of the experiment. Error bars denote the standard error in the mean of several measurements. }
\label{fig:nASumz}
\end{figure}

In contrast, neither~$n_0$ nor~$A$ show systematic variation with surface temperature (Fig.~\ref{fig:nASumz}a and b respectively):~$n_0$ fluctuates around 2 collisions, and~$A$ fluctuates around 2 pC/collision. Returning to the phenomenological model in Eq.~(\ref{eq:qn}), we interpret~$n_0$ as a characteristic timescale for charge exchange, and~$A$ as a slow background increase in the total charge. The fact that~$A$ does not vary with temperature suggests that this linear term originates in a systematic feature of the experiment. Similarly, the fact that~$n_0$ does not show systematic variation with temperature suggests that although the temperature affects the total capacity to exchange charge (Fig.~\ref{fig:qSumz}), it does not affect the timescale (and, by implication, the mechanism) by which this charge is exchanged.  

\section{Model for temperature dependence of charge exchange}
\label{sec:model}
\begin{figure*}
\centering
\includegraphics[width = 1.8\columnwidth]{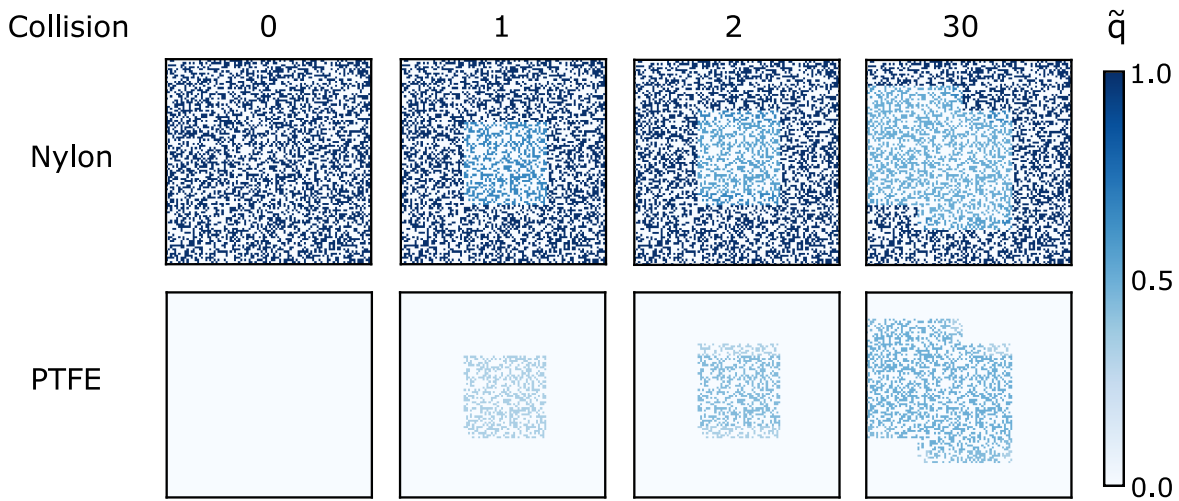}
\caption{Simulated example of the charge transfer between two surfaces. Darker colours indicate more charge (equivalently, surface water in this model). At the beginning of the simulation, all the surface water is on the nylon. As the number of collisions increases (indicated by the numbers on top), charge on the nylon is transferred to the PTFE. }
\label{fig:ChargingSequence}
\end{figure*}

In order to explain our experimental results, we develop a simple computational model for contact charging. We base our model on the charging mechanisms proposed in Refs.~\cite{wiles2004effects,thomas2008patterns,mccarty2008electrostatic,lee17,lee2018collisional}, where the transfer of charge is associated with the collision of nanoscale dry and wet patches on the surfaces. Dry (wet) here refers to the absence (presence) of molecularly thin surface layers of water, which are ubiquitous under ambient conditions. During such a collision, adsorbed OH$^-$ ions can be transferred from a wet area on one surface to a dry area on another surface~\cite{yatsuzuka1996electrification,zimmermann2001electrokinetic,kudin2008water}. Depending on the number of dry and wet patches, which depends on the hydrophobicity of the surface~\cite{ewing2006ambient,kudin2008water,xu2010graphene}, there may be a net charge transfer as one surface transfers its wet patches to another surface~\cite{yatsuzuka1994electrification,yatsuzuka1996electrification,ravelo2011demonstration,burgo2016water}. 

We conduct a Monte Carlo simulation of the two insulator surfaces in order to model this charge transfer process. These surfaces are partitioned into wet and dry patches. As an initial simplification, we treat the hydrophobic PTFE surface as if it did not have any wet patches (see Appendix D for supporting data). By contrast, the nylon surface is initially covered with a random distribution of wet patches (see Appendix E for details). Here, we use the parameter~$p_\mathrm{wet}$ (which increases the proportion of wet to dry patches) as the control variable for the amount of surface water. We assume that the amount of negative charge (OH$^-$ ions) available for transfer is directly proportional to the amount of water on the insulator surface. An example of the simulated initial distribution of surface water (and thus negative surface charge) on the nylon and PTFE is shown in Fig.~\ref{fig:ChargingSequence}.  

Given the initial distribution of charges on the nylon and PTFE pieces, we then implement collisions by transferring a fraction $f_t$ of the charge on a wet patch to a corresponding dry patch on the PTFE.
In addition, we stochastically translate the area of contact between the nylon and PTFE surfaces for each collision. This mimics our experimental conditions, where the areas of contact can vary from collision to collision because of inadvertent, small amounts of drift in the linear actuator. To implement this in the simulations, we  used the following scheme. We defined a square with side length $w$, called the contact area, and using a random walk, we displaced the contact area a distance $d$ in a random direction each collision. During each collision, only the elements within the contact area were allowed to exchange charge. In the simulation, this produces a slowly increasing background that can be approximated by a linear term~$\tilde{A}$ in the charge accumulation, similar to what we see in our experiments. Examples of the evolution of the surface charge distribution over several collisions are shown in Fig.~\ref{fig:ChargingSequence}. 

In order to calibrate the simulations to the effect of temperature on the nylon and PTFE surfaces, we measured the mass of water adsorbed by a piece of nylon at different temperatures using thermogravimetric analysis (TA Instruments Q600 SDT). We assume that the mass of surface adsorbed water decreases proportionally to the bulk water content, since the mass of surface adsorbed water is too small to measure directly. We heat a piece of nylon ($l\times w\times d = 3.8\times3.8\times0.5$ mm$^3$, mass = 6 mg) at a rate of 20~$^\circ$C/minute, until it reaches a specified temperature between 30 and 80~$^\circ$C. The temperature is then held constant, and the mass of the sample is measured as a function of time. Data was taken at atmospheric pressure, under inert nitrogen (flow rate 100mL/minute). Prior work has shown that the onset decomposition temperature of nylon is over 400$^\circ$C~\cite{carrizales2008thermal}, suggesting that mass loss over the range of temperatures that we probe can be attributed to the evaporation of adsorbed water. We confirm that the mass loss we observe is due only to the evaporation of water by comparing the mass loss curves from a piece of nylon soaked in water overnight, and a piece of nylon dried in a vacuum oven (see Appendix C). 

\begin{figure}
\centering
\includegraphics[width = \columnwidth]{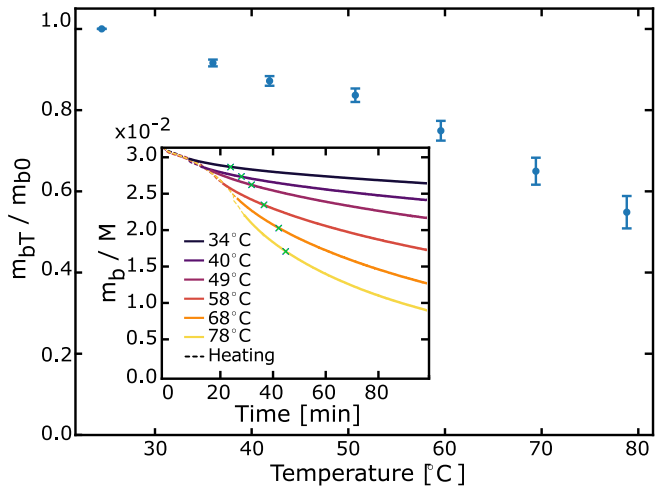}
\caption{Temperature variation of the water content of a nylon sample. 
Plot of mass fraction of remaining bulk water~$m_{bT}/m_{b0}$ as a function of temperature, 15 minutes after being held at constant temperature. The water content of the nylon decreases over time, at a rate which is proportional to temperature. The error bars denote bulk water proportions after waiting $15\pm 5$ minutes after the sample has reached the target temperature.
Inset: Plot of bulk water content~$m_b$ normalised by initial sample mass~$M$ as a function of time. Different colours indicate different temperatures (see figure legend). Note that the data are shown as solid lines only after the target temperature has been reached. Blue crosses indicate the times at which~$m_{bT}$ was measured (15 minutes after target temperature was reached).}
\label{fig:TGA}
\end{figure}

We first measure the total mass percentage of water adsorbed by a piece of nylon stored in ambient conditions by heating the sample to 100$^\circ$C and waiting for the mass of the sample to equilibrate (see Appendix C for data). Denoting the total initial mass of the nylon sample to be~$M$, including a mass of~$m_\mathrm{b0}$ adsorbed water, we measure the initial fractional water content of our nylon samples to be~$m_\mathrm{b0}/M = 0.0316$. We then heat samples to the temperatures we use in the experiment, measuring the fractional mass of water~$m_\mathrm{b}/M$ as a function of time at each temperature (inset to Fig.~\ref{fig:TGA}). We mirror the conditions of the charging experiment by measuring the percentage of remaining adsorbed water as a fraction of the original mass of water,~$m_\mathrm{bT}/m_\mathrm{b0}$ after 15 minutes at the target temperature. Data are shown in the main panel of Fig.~\ref{fig:TGA}. We find that the mass of water remaining in the sample decreases with temperature, and that heating to just 80~$^\circ$C for 15 minutes evaporates 40\% of the water content of a nylon sample, while there is no measurable effect for the PTFE (see Fig.~\ref{fig:tga_teflon} and Appendix D for data on the water content of a PTFE sample).

\begin{figure}
\centering
\includegraphics[width = \columnwidth]{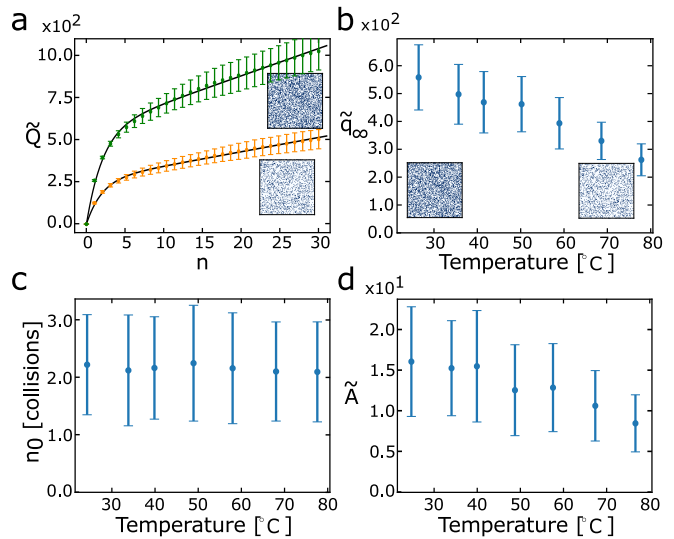}
\caption{Variation of simulation results with temperature, as captured by the proportion of wet to dry patches on the surface~$p_\mathrm{wet}$. We use the results from Fig.~\ref{fig:TGA} to calibrate~$p_\mathrm{wet}$ to the temperature of the nylon. (a) Averaged charging data for dimensionless charge on the PTFE~$\tilde Q$ as a function of number of collisions~$n$. Green (orange) curves were simulated for~$p_\mathrm{wet}=0.60$ ($p_\mathrm{wet}=0.29$). The curves are fit to Eq.~(\ref{eq:qn}). An example initial charge distribution on the simulated nylon is shown next to the corresponding curve. (b) Simulated variation of the dimensionless fit parameter~$\tilde{q}_\infty$ with temperature. $\tilde{q}_\infty$ decreases with temperature. Images along the bottom axis indicate the initial charge distribution on the simulated nylon for different values of temperature ($p_\mathrm{wet}$). (c) Simulated variation of~${n}_0$ with temperature. $n_0$ fluctuates with temperature. (d) Simulated variation of the dimensionless fit parameter~$\tilde{A}$ with temperature. $\tilde A$ shows large fluctuations with temperature. Error bars in (a), (b), (c), and (d) denote the standard deviation in the distribution of 200 simulations per temperature. For these simulations, $f_t = 0.27$ and $w/d = 8$.}
\label{fig:BigSimSum}
\end{figure}

We use these results to proportionally decrease the proportion of wet patches on the simulated nylon by setting ~$p_\mathrm{wet}(T)~=~p_0\times~m_{bT}/m_{b0}$, $p_0 = 0.6$ (see Appendix E). This reproduces the effect of temperature on the transfer of charge. Fig.~\ref{fig:BigSimSum}a shows simulated data for the total (dimensionless) charge~$\tilde{Q}$ on the nylon as a function of collision number~$n$, for two different temperatures (green, orange data points). The initial distribution of wet patches on the surface of the nylon is shown for each temperature (inset). Qualitatively, we reproduce the shape of the experimental results shown in Fig.~\ref{fig:Apparatus}b, and so we fit the simulation data also to Eq.~(\ref{eq:qn}) (black lines). The variation of the fit parameters with temperature is shown in Figs.~\ref{fig:BigSimSum}b-d. In particular, we find that~$\tilde{q}_\infty$ decreases linearly with temperature, in proportion to the experimental results in Fig.~\ref{fig:qSumz} (a decrease by a factor of two over the same range of temperatures), suggesting that the simulations have captured the effect of temperature in the experiments. In addition, the simulated value of~$n_0$ fluctuates just below 2 collisions, again in good agreement with the experiments. 


We note that in the simulation there is a decreasing trend in~$\tilde{A}$ with temperature, reflecting the decreasing proportions of water on the surfaces. We do not see this trend in the experimental data in Fig. 4b; however, it is possible that larger temperatures increase the drift, which could compensate at least for part of the loss of surface water.



Our model suggests a means to reconcile previously contradictory results. While some experiments have observed an exponential increase of charge during contact charging experiments~\cite{collins2018simultaneous,liu2013contact,greason1999investigation,apodaca2010contact, matsusaka2010triboelectric}, other measurements on single grains have observed a linear, rather than exponential, charging when the collision area is randomised~\cite{lee2018collisional}. In our model, most charge is exchanged after a few collisions, with a slow, stochastic background motion of the contact area. After 2-3 collisions, the two surfaces have exchanged most of their wet patches within the contact area, leading to an exponential saturation. This local saturation is offset on further collisions by the gradual motion of the contact area, making new dry and wet patches accessible. Since the new wet patches are revealed slowly, the result is that the charge increases much more slowly than on initial contact. We propose that this separation of timescales between collisions and the motion of the contact area leads to the observed exponential charging with linear term that we observe. If the contact area motion is much faster, i.e., a new set of surface patches comes into contact every collision, then the surfaces will not locally equilibrate and thus only the linear term will be observed. In contrast, we recover purely exponential charging if the timescale for contact area motion becomes very large. Within this framework, then, the interplay of local saturation and stochastic contact motion is able to explain the variety of observed charging trends.

\section{Conclusions}
We have presented an experiment and simulation to study collisional charging between a pair of macroscopic dielectrics.
Although the overall nature of contact charging is stochastic, and depends on the microscopic details of the two surfaces that come into contact, our results nevertheless show that statistically significant trends can be extracted from macroscopic experiments. In particular, we demonstrate that the magnitude of charge transfer between a piece of nylon and a piece of PTFE decreases as their surface temperature is increased, while the charge transfer trends remain constant. The agreement between our experiments and simulations, combined with TGA measurements, suggest a contact charging mechanism for nonionic insulators that relies on the presence of surface water. Charge is carried by OH$^-$ ions that are transferred at the sites where a wet patch on one of the surfaces collides with a dry patch on the other surface.  Evaporating the surface water via a modest increase in temperature thus leads to a decrease in the overall magnitude of charge transfer, while not affecting the initial rate of exponential charge saturation.

Our experimental method can be used straightforwardly for measurements on other materials, in a wide range of different environments. We anticipate that similar experiments will continue to inform an understanding of the basic mechanism of collisional charge transfer. In addition, we envision that such macroscopic, statistical measurements will shed light on processes in tribocharging that are especially relevant to industrial applications. 

\section*{Acknowledgments}

We thank Grayson Jackson for his assistance with the TGA measurements, and Victor Lee for his work on the early stages of the experiment. We thank Kieran Murphy, Leah Roth, Abhinendra Singh, and Nicole James for insightful discussions. This research was supported by the National Science Foundation through grant DMR-1810390. I.A.H. acknowledges primary support from the Chicago MRSEC, funded by the NSF through grant DMR-1420709.

\section*{Appendix A: Sample preparation}

Samples of Nylon 6,6 (McMaster Carr) were cut to size, then rinsed with toluene, ethanol, and then DI water. Samples of PTFE (McMaster Carr) were cut to size, then rinsed with toluene, isopropanol, ethanol, and DI water. For both samples, excess water was dried with high-purity N$_2$ gas and then placed in a vacuum chamber with an oil-free pump for several hours. The plastics were then transferred into the apparatus and were left to discharge under ambient conditions for 5 or more hours. Using heating pads attached to the mounts for the nylon and PTFE, we then heated the plastics to the desired temperature with PID controllers, attached to thermistors on the underside of the plastics.

\section*{Appendix B: Temperature calibration}

\begin{figure}
\centering
\includegraphics[width = 0.9\columnwidth]{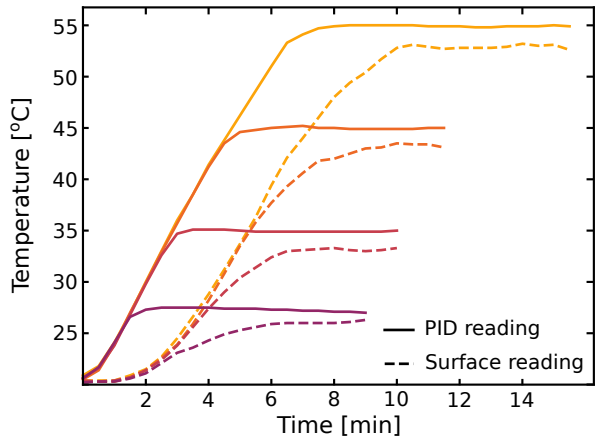}
\caption{Temperature as a function of time for four different target temperatures, measured either with the PID controller (solid lines, thermistor placed on the underside of the sample), or with a thermistor placed on the sample surface (dashed lines). In the experiment and in the TGA measurements, we wait 15 minutes for the sample temperature to equilibrate in order to avoid the transients shown here. Data in the main text are plotted as a function of the surface temperature.}
\label{fig:tcalibration}
\end{figure}

In order to calibrate our temperature sensors, we ran tests outside of the apparatus with the PID controllers set to heat to a specified temperature, while we also monitored the surface temperature of the plastics with a separate thermistor placed on the sample surface. Our results for PTFE are shown in Fig.~\ref{fig:tcalibration}, on the basis of which we wait 15 minutes for the temperature of the samples to equilibrate (and thus take readings after 15 minutes at constant temperature in the TGA measurements). The results are identical for the nylon samples. We use the actual surface temperature for the data shown in the main text. 

\section*{Appendix C: TGA measurements for dry and wet nylon}

\begin{figure}
\centering
\includegraphics[width = 0.88\columnwidth]{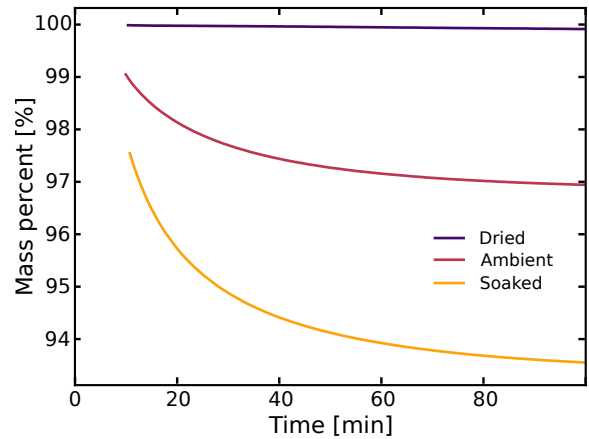}
\caption{TGA measurements of mass loss as function of time at 100$^\circ$C. Data are only shown after the temperature has equilibrated. Curves correspond to three different preparation methods: 1) nylon dried in a vacuum oven for 2 hours at 100-140$^\circ$C, then held at 10 hours under vacuum, as temperature slowly relaxed to room (purple) 2) nylon stored at ambient conditions (magenta), and 3) nylon soaked in DI water for 12 hours (yellow). }
\label{fig:tga_water}
\end{figure}

In order to verify that the mass loss we measure is due to the evaporation of bulk water, we performed control measurements with three nylon samples ($l\times w\times d = 3.8\times3.8\times0.5$ mm$^3$, mass = 6 mg). Our results are plotted in Fig.~\ref{fig:tga_water} . One of the samples was soaked in DI water for 12 hours (yellow curve). The other nylon sample was dried in a vacuum oven for 2 hours at 100-140$^\circ$C (uncertainty due to the vacuum oven), then held at 10 hours under vacuum (15 psi) in the same oven (purple curve). These were also compared to a sample stored at ambient conditions (same as the other samples we test in the main text, magenta curve). Samples were heated at a rate of 10~$^\circ$C/minute, until a temperature of 100$^\circ$C was reached. The temperature was then held constant, and the mass of the sample measured as a function of time. Data was taken at atmospheric pressure, under inert nitrogen (flow rate 100mL/minute).

We find that the sample that was dried shows very little mass loss compared to the sample stored at ambient conditions, and that the mass loss is greater (by a factor of 2) from the soaked nylon. We thus conclude that the temperature-dependent mass loss we measure can be attributed to the loss of bulk water, within the accuracy of our experiment.

\section*{Appendix D: TGA measurements for PTFE}

\begin{figure}
\centering
\includegraphics[width = 0.92\columnwidth]{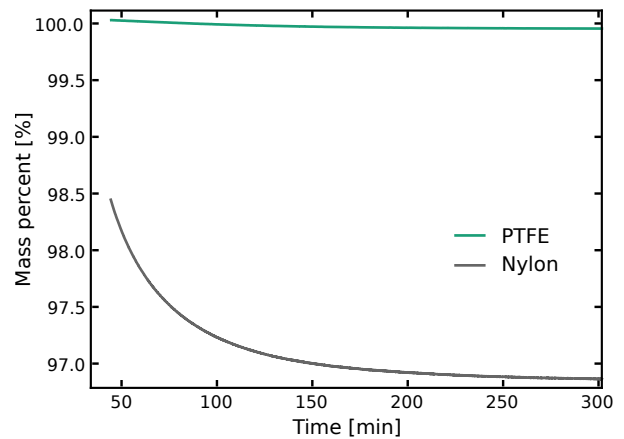}
\caption{TGA measurements of mass loss as function of time at 100$^\circ$C, for a sample of nylon (grey) and a sample of PTFE (green). Data are shown only after the ramp time of the TGA. Over five hours, the nylon loses~$\sim$ 3\% of its mass, while the PTFE loses only~$\sim$ 0.07\% of its mass. }
\label{fig:tga_teflon}
\end{figure}

In order to justify our approximation in the simulations that the PTFE begins with no wet patches, we measured the mass loss of a PTFE sample ($l\times w\times d = 3.8\times3.8\times0.5$ mm, mass = 18 mg) using TGA. Samples (stored in ambient conditions) were heated at a rate of 2~$^\circ$C/minute, until a temperature of 100$^\circ$C was reached. The temperature was then held constant, and the mass of the sample measured as a function of time. Data were taken at atmospheric pressure, under inert nitrogen (flow rate 100mL/minute). Data are shown in Fig.~\ref{fig:tga_teflon}. The mass loss from a nylon sample stored in the same conditions is also plotted for comparison. Over the course of 5 hours, the mass of the PTFE equilibrated to 99.93\% of its original mass (in contrast to~$\sim$ 97\% for the nylon sample). Our results suggest that the mass of water adsorbed by the PTFE is~$\sim$50 times smaller than the mass of water adsorbed by the nylon.  

\section*{Appendix E: Simulation protocol}
We stochastically generate the wet patches on the nylon. We initialise a 100$\times$100 matrix~$N$, representing the nylon surface, whose entries are either 1 or 0, with probability~$p_\mathrm{wet}$ of generating a 1 in any given entry. The nonzero entries of the matrices serve as  wet patches. Typical values of~$p_\mathrm{wet}$ in our simulations were between 0.3 and 0.6. We choose the proportion of wet patches on room temperature nylon as~$p_0=0.6$. Since we are only interested in the proportionality in $p_{\mathrm{wet}}$ across different temperatures as measured by the TGA, the value of $p_0$ is not important. As described in the main text, we simulate the effect of increasing temperature (and thus decreasing surface water) by decreasing~$p_\mathrm{wet}$ in proportion to our TGA measurements (Fig.~\ref{fig:TGA}):~$p_\mathrm{wet}(T)= p_0\times m_{bT}/m_{b0}$. 

We stochastically translate the area of contact between two surfaces in order to simulate the drift of the actuator, or the general change in charging surfaces over repeated contacts. For any given collision, we model the contact area as a square of side length~$w$. Surfaces outside of this contact area do not exchange charge. Over the course of the simulated experiment, this contact area is displaced over the nylon matrix as a random walk with a displacement~$d$ each collision, with the direction of the displacement also chosen randomly from the four cardinal directions (for instance~$d=6$ means that the nylon contact area shifts by 6 matrix elements in the $\pm x$ or $\pm y$ direction for each collision). For the results shown in the main text, $d=5$ and~$w=40$. For a discussion of how the results change for different parameter choices, see Appendix F.

After the area of contact has been stochastically translated in this way,  we implement our charge transfer rules: we compare $T_{ij}$ with $N_{ij}$ only within the designated contact area. If the local charge on the PTFE is smaller than the corresponding local charge on the nylon, they exchange a fraction $f_t$ of the charge difference between them. Formally, within the contact area, the local charge on the PTFE after collision~$T^\prime_{ij}$ is given by

\begin{align}
T^\prime_{ij}=T_{ij} + f_t(N_{ij} - T_{ij})
\label{eq:teflonex}
\end{align}

To conserve charge, the corresponding local charge on the nylon after collision~$N^\prime_{ij}$ is 

\begin{align}
N^\prime_{ij}=N_{ij} - f_t(N_{ij} - T_{ij})
\label{eq:nylonex}
\end{align}

The total charge on the PTFE (as plotted in Fig.~\ref{fig:BigSimSum}a) is then~$\tilde{Q}_T = -\sum_{ab} T_{ij}$. By conservation of charge, the total charge on the nylon~$\tilde{Q}=-\tilde{Q}_T$. 

As shown in Fig.~\ref{fig:temporary_parameter_exploration}(c), $f_t$ controls the rate of the exponential charge saturation $n_0$ in our simulation: as $f_t$ decreases, $n_0$ increases.  In order to match our simulation (Fig.~\ref{fig:BigSimSum}) to our experiments (Fig.~\ref{fig:nASumz}), we set $f_t$ to 0.27. We note that in the experiment, several different physical mechanisms could control $n_0$. In the simulation, we summarise the combined effect of these mechanisms with the single variable~$f_t$.

\section*{Appendix F: Robustness of model results}

\begin{figure}
\centering
\includegraphics[width = 0.92\columnwidth]{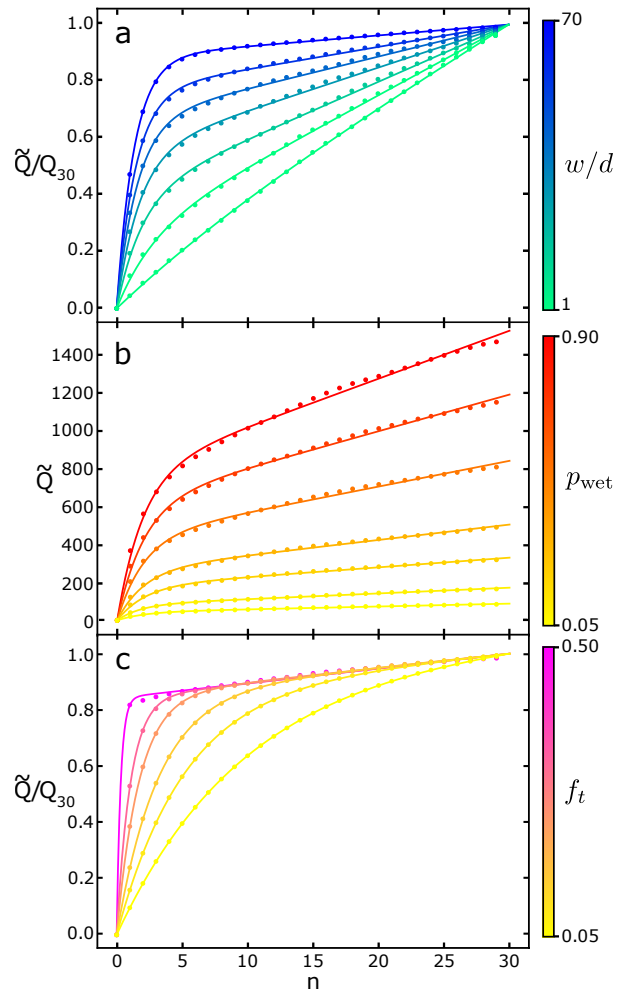}
\caption{Simulated charge~$\tilde{Q}$ as a function of number of collisions~$n$, shown for varying values of the 3 main parameters of the simulation: (a) the ratio of contact area width to displacement per collision,~$w/d$, (b) the proportion of wet patches to dry patches~$p_\mathrm{wet}$, and (c) the fraction of charge transferred per collision from a wet patch to a dry patch~$f_t$. In order to compare qualitative trends, the charge sequences in (a) and (c) are normalised by~$Q_{30}$, the charge at~$n=30$. In (a),~$p_\mathrm{wet} = 0.4,$ $f_t = 0.27$. In (b),~$w/d = 8$, $f_t = 0.27$. In (c),~$p_\mathrm{wet} = 0.4,$ $w/d = 50$. }
\label{fig:temporary_parameter_exploration}
\end{figure}

In order to test the robustness of the results that we present in Sec.~\ref{sec:model} of the main text, we reproduce our simulation model results across a wide range of parameters. In Fig.~\ref{fig:temporary_parameter_exploration}, we vary the displacement per collision~$d$, the proportion of wet patches on the nylon $p_\mathrm{wet}$, and the fraction of charge transferred per collision $f_t$. In order isolate the effect of each parameter, the (random) motion of the contact area was generated for one charge sequence, then used for all other parameter choices.  

Fig.~\ref{fig:temporary_parameter_exploration}(a) shows normalised charge time-series for varying values of~$w/d$, the ratio of the width of the contact area~$w$ and the displacement of the contact area per collision~$d$. For small~$w/d$, the displacement of the contact area and the width of the contact area are of comparable size, leading to small overlaps in the areas that exchange charge from collision to collision. This lack of self-overlap leads to a lack of saturation in the underlying surface, and thus to a linear increase in charge with collision number. As~$w/d$ is increased, the overlap of the contact area from collision to collision increases. Correspondingly, the charge time-sequences become increasingly exponential. 

Fig.~\ref{fig:temporary_parameter_exploration}(b) shows charge time-series for different values of~$p_\mathrm{wet}$, the proportion of wet patches on the nylon matrix. We note that increasing~$p_\mathrm{wet}$ changes only the magnitude of the final charge (~$q_\infty$) and not the timescale of saturation or the functional form of the charge increase. In contrast, Fig.~\ref{fig:temporary_parameter_exploration}(c) shows that the timescale of the exponential charge saturation is controlled by~$f_t$, the fraction of charge transferred each collision. 

\section*{Appendix G: Surface Characterization}

\begin{figure}
\centering
\includegraphics[width = 0.92\columnwidth]{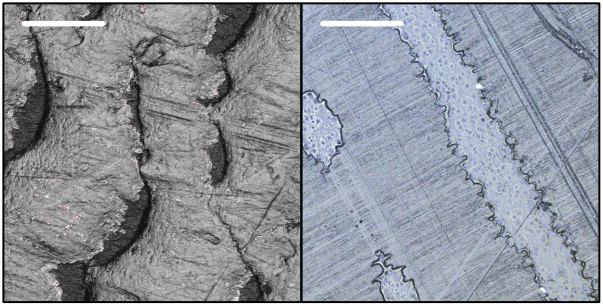}
\caption{Surface scans of PTFE (left) and nylon (right). The scale bars are 100~$\mu$m. The vertical minimum to maximum distances are 28~$\mu$m for the PTFE, 0.5~$\mu$m for the nylon sample.}
\label{fig:SurfacePics}
\end{figure}

Images of the microscopic surface structure of the samples (see Fig.~\ref{fig:SurfacePics}) were obtained using an Olympus LEXT OLS5000 surface scanner.

\end{document}